\documentclass{aa}

\usepackage{graphicx}
\usepackage{pjournal}
\usepackage{mamd}
\usepackage{natbib}
\usepackage{amssymb}

\begin{document}

\title{Power spectrum of the cosmic infrared background at 60 and 100 \um with IRAS}

\author{M.-A. Miville-Desch\^enes\inst{1}\and G. Lagache\inst{2} \and J.-L. Puget\inst{2}}
\institute{Laboratoire de radioastronomie, \'Ecole Normale Sup\'erieure, 24 rue Lhomond, 75005, Paris, France
\and Institut d'Astrophysique Spatiale, Universit\'e Paris-Sud, B\^at. 121, 91405, Orsay, France.}

\offprints{Marc-Antoine Miville-Desch\^enes}
\mail{miville@lra.ens.fr}
\date{\today}
\abstract{
Based on a power spectrum analysis of the IRAS ISSA maps, we present 
the first detection of the Cosmic far-Infrared Background (CIB) fluctuations at 60 and 100 \ump. 
The power spectrum of 12 low cirrus emission regions is
characterized by a power excess at spatial frequencies higher than
$k \sim 0.02$ arcmin$^{-1}$.
Most of this excess is due to noise and to nearby point sources with a flux stronger than 1 Jy.
But we show that when these contributions are carefully removed, there is still a power excess that is
the signature of the CIB fluctuations.
The power spectrum of the CIB at 60 and 100 \um is 
compatible with a Poissonian distribution,
at spatial frequencies between 0.025 and 0.2 arcmin$^{-1}$.
The fluctuation level is $\sim 1.6\times10^{3}$ Jy$^2$/sr and 
$\sim 5.8\times10^{3}$ Jy$^2$/sr at 60 and 100 \um respectively.
The levels of the fluctuations are used in a larger framework, with other observationnal
data, to constrain the evolution of IR galaxies (Lagache et al. 2002). The detections reported here,
coupled with the level of the fluctuations at 170 $\mu$m,
give strong constraints on the evolution of the IR luminosity function.
The combined results at 60, 100 and 170 $\mu$m for the CIB and its fluctuations
allows, on the CIB at 60 $\mu$m, to put a firm upper limit of 0.27 MJy/sr
and to give an estimate of 0.18 MJy/sr.
}
\titlerunning{Power spectrum of the CIB at 60 and 100 \um}

\maketitle

\section{Introduction} 
The Cosmic Far-Infrared Background (CIB) was detected 
by \cite{puget96}, exploiting COBE-FIRAS data, and has now been firmly
established over a range of wavelengths (e.g. \cite{dwek98,gispert2000,hauser2001}).
The intensity is quite high with respect 
to predictions based on evolutionary models of star formation in galaxy
populations inferred from optical data. Source counts obtained with SCUBA
at 850 $\mu$m and ISO at 170 and 15 $\mu$m have partly
resolved the CIB and shown a strong cosmological evolution. 
In the future, far-IR and sub-millimeter telescopes
from ground and space will perform deep surveys over small
areas, aimed at resolving a substantial fraction of the CIB and to shed
light on the number density, luminosity and spectral evolution of the
infrared galaxy populations.  However, investigation of the clustering of these
populations requires surveys over much larger areas.  
One way to tackle the limitation on the number of detected galaxies
per field is to search for CIB fluctuations.
So far, CIB fluctuations has only been observed at 170 $\mu$m
in the FIRBACK fields \cite[]{lagache2000} and at 90 and 170 $\mu$m in the Lockman hole
\cite[]{matsuhara2000} surveys. These detections 
were probably dominated by the Poissonian contribution. 
By analyzing larger FIRBACK fields, power spectra seem to reveal correlated fluctuations, well 
above the cirrus contribution \cite[]{puget2002}.
These results are currently under further investigation.
Above $170\,\mu$m, only SCUBA
observations might currently be able to give information on the
fluctuations. However, no significant CIB correlations have been detected in
the SCUBA maps \cite[]{peacock2000}.
The CIB anisotropies are mainly contributed by moderate to
high redshift star-forming galaxies, whose clustering properties and
evolutionary histories are currently unknown.
Since the clustering strength depends on the bias at the relevant redshift,
observing the CIB correlated fluctuations will provide valuable informations
on bulge and elliptical formation, as well as potentially QSOs,
thereby providing clues on the physical relations between dark matter and
starburst galaxies. 
\\
In this paper we present the results of a power spectrum analysis of the
IRAS 60 and 100 \um emission of 12 regions in the sky with very low interstellar emission.
All these regions are characterized by an excess of power at high spatial frequencies
with respect to the interstellar emission.
We will show that this excess is not of instrumental origin and may be attributed to the CIB.
The discovery of the CIB fluctuations at 60 and 100 $\mu$m could give
constraints on the number counts below the IRAS point source detection limit.
However, instead of extrapolating the number counts at 60 and 100 $\mu$m
using the level of the fluctuations, which is of limited cosmological
interest, we prefer to use the two detections in a more general 
framework of the modelisation of the IR galaxy evolution that combines 
all existing number counts, redshift distributions, and observations of the CIB and its
fluctuations, in the whole IR and submm range \cite{lagache2002}.
\\
The paper is organized as follow.
After a presentation (\S~\ref{data}) of the data used for this
analysis we describe in \S~\ref{powerspectrum} the power spectrum
at 60 and 100 \um of the 12 fields selected. In \S~\ref{cib} we
present the method used to separate the Galactic and extra-Galactic 
contributions to the power spectrum and we discuss our results in \S~\ref{discussion}.


\section{Data - the ISSA maps}

\label{data}

\subsection{The IRAS Sky Survey Atlas}

The IRAS mission was originally designed to measure
point sources but it also provided spectacular images of the 
Galactic dust diffuse emission. These images, 
constituted of $500\times500$ pixels with a pixel size of
1.5' (i.e. the size of the field is $12.5^\circ \times 12.5^\circ$),
are gathered in the IRAS Sky Survey Atlas (ISSA).
Each ISSA map is the result of the combination of up to three
individual maps (named for HCON Hours CONfirmation).
The HCON images were constructed from different
observations of the same region separated by several months. 
The angular resolution of the 60 and 100 \um band images are
respectively $1.5'\times4.7'$ and $3.0'\times5.0'$.
To study the power spectrum of the CIB emission at 60 and 100 \ump, we have
worked on 12 ISSA maps with particularly low cirrus emission. 

\subsection{Calibration of the ISSA maps}

Special care has been
taken to have a consistent diffuse emission calibration through the Atlas.
But, as the IRAS mission was designed to provide absolute
photometry only for point sources,
the ISSA images give only relative photometry and cannot be used
to determine the absolute surface brightness for diffuse emission.
Based on a comparison with the DIRBE data, it has been shown
\cite[]{wheelock93,schlegel98} that, at large scale ($>$ a few degrees), the
amplitude of the fluctuations are overestimated in the ISSA maps by a
factor 1.15 at 60 \um and 1.39 at 100 \ump. We have thus applied these
factors to our maps prior to the analysis. 

There is also an uncertainty on the zero level of the ISSA maps, 
for which the zodiacal emission has been subtracted \cite[]{wheelock93}. 
This uncertainty is dominated at 60 and 100 \um by an imperfect knowledge of the detector offsets
and of the zodiacal emission. As we are looking at fluctuations of the signal, 
a global additive offset has no impact on the power spectrum.
On the other hand, an imperfect zodiacal emission correction applied to the ISSA maps
will have an impact on the large scale structure of the maps and then on
the power spectrum (for low $k$ values). To restore appropriately the
large scale structure of the selected fields we have compared each ISSA map
with the DIRBE data (for which the zodiacal emission correction was better done).
The ISSA maps, multiplied by the appropriate gain value (0.87 at 60 \um and 0.72 at 100 \ump) 
were convolved by the DIRBE beam and then subtracted from the DIRBE data.
This offset map is then added to the ISSA map. 
Note that we do not use the \cite{schlegel98} IRAS rescaled maps, constructed
in a similar manner, for which it is impossible to recover the associated instrumental noise.

\subsection{Fields with low cirrus emission}

In this paper we present a power spectrum analysis of 
twelve high latitude fields. These fields were selected on the 
basis of their low cirrus emission (mean brightness $\sim 1$ MJy/sr at 100 \ump)
but also on their redundancy;  we have selected only
fields for which each sky position has been observed at least twice
in order to be able to estimate the contribution of the noise to the power spectrum.
The typical stripping of the ISSA maps is very efficiently removed from
the signal power spectrum by the noise estimate procedure using the difference
between observations of the different HCON as will be shown in \S~3.3 and Fig. 2.
\\
The 60 and 100 \um ISSA maps of the twelve selected fields, gain and offset corrected,
are shown in Figs~\ref{pscib1} to \ref{pscib_fin} and their central coordinates are 
gathered in Table~\ref{coordinate_table}.
Five fields are located in the southern hemisphere, 
most of them in the neighborhood of the Marano field \cite[]{marano88}, and 
seven are spread over the northern hemisphere (ISSA map number 376 contains the Lockman Hole). 

\begin{table}
\begin{tabular}{lcccc}\hline
ISSA & $\alpha_{2000}$ & $\delta_{2000}$ & l  & b \\ \hline
47 & 4h 1m 24.8s & -49$^\circ$ 51' 41.5'' & 258.32$^\circ$ & -47.41$^\circ$\\
66 & 23h  2m 53.8s & -49$^\circ$ 43'    50" & 338.21$^\circ$ & -59.30$^\circ$ \\
69 &  1h 46m  9.0s & -39$^\circ$ 45'     1" & 264.42$^\circ$ & -73.02$^\circ$ \\
71 &  3h 29m 49.5s & -39$^\circ$ 49'    46" & 244.45$^\circ$ & -54.96$^\circ$ \\
97 &  1h 34m 18.6s & -29$^\circ$ 44'    39" & 230.96$^\circ$ & -80.22$^\circ$ \\
322 & 13h  4m 23.3s &  29$^\circ$ 43'    55" &  76.14$^\circ$ &  86.14$^\circ$ \\
323 & 13h 50m 16.1s &  29$^\circ$ 45'     9" &  47.86$^\circ$ &  76.81$^\circ$ \\
348 &  9h 35m  7.0s &  39$^\circ$ 46'    35" & 182.59$^\circ$ &  47.71$^\circ$ \\
349 & 10h 26m 56.1s &  39$^\circ$ 44'    41" & 180.70$^\circ$ &  57.59$^\circ$ \\
356 & 16h 29m 42.1s &  39$^\circ$ 53'    31" &  63.41$^\circ$ &  43.50$^\circ$ \\
375 & 10h  3m 12.9s &  49$^\circ$ 45'    28" & 166.14$^\circ$ &  50.80$^\circ$ \\
376 & 11h  2m 53.8s &  49$^\circ$ 43'    50" & 158.21$^\circ$ &  59.30$^\circ$ \\\hline
\end{tabular}
\caption{\label{coordinate_table} Central pixel coordinates (ecliptic and Galactic)
of the 12 ISSA maps selected for our analysis.}
\end{table}


\section{Power spectrum of the 60 and 100 \um sky}

\label{powerspectrum}

\subsection{Power spectrum computation}

For a given image, the power spectrum $P(k)$ is
the absolute value of the Fourier Transform (we use
the Fast Fourier Transform (FFT) function of IDL)
of the image, averaged over constant values of 
$k=\sqrt{k_x^2+k_y^2}$. To make sure that map boundaries are not
contaminating the computed power spectrum, we apodize the image with a
cosine function before computing the Fourier Transform. This prevents
the introduction of strong discontinuities when the image is periodized
in the FFT algorithm (see Appendix~\ref{cross_fft}).

\begin{figure*}[!ht]
\hspace{-0.7cm}
\includegraphics[width=\linewidth]{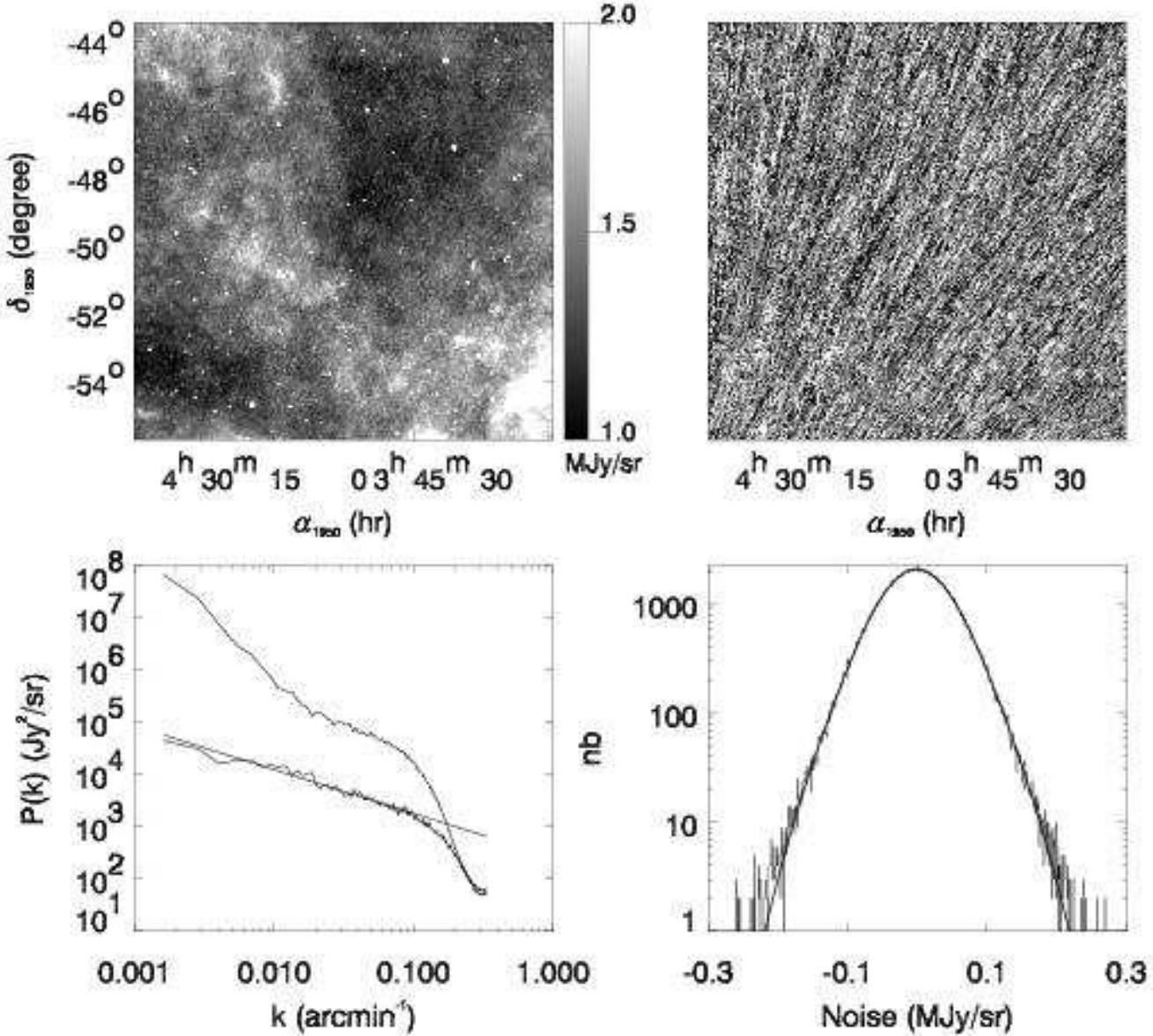}
\caption{\label{noise_ps} {\bf Top left:} The 100 \um ISSA map number 47.
{\bf Top right:} Noise map of the 100 \um map on the left. This noise map
has been obtained by subtracting two individual HCON images that were
used to build the ISSA map (see text for details). {\bf Bottom left:} Power spectrum of the ISSA
map (upper curve) and of the noise map. The power spectrum of the
noise map has been fitted by a power law $\propto k^{-0.83}$ for scales $k
\leq 0.1$ arcmin$^{-1}$. Above k=0.1 arcmin$^{-1}$, the drop of the power spectrum is
due to the instrumental function (see \S~\ref{secPSF}). {\bf Bottom right:} Histogram of the noise values, fitted
by a sum of two Gaussian functions (Amplitude$_1$ = 1376, $\sigma_1$ = 0.04 MJy/sr; 
Amplitude$_2$ = 711, $\sigma_2$ = 0.06 MJy/sr).}
\end{figure*}

The power spectrum of the 100 \um ISSA map of one of our fields is shown in Fig.~\ref{noise_ps}.
Since the zodiacal emission has been subtracted in the ISSA maps, 
there are four main contributions to the power 
spectrum at these wavelengths:
the cirrus emission, point sources, 
the CIB (resolved and unresolved) and the noise.
Therefore, and if the noise and the signal are not correlated, 
the power spectrum $P(k)$ 
can be expressed in the following manner:
\begin{equation}
\label{pk}
P(k) = \gamma(k) \left[P_{\rm dust}(k) + P_{\rm source}(k) + P_{\rm cib}(k) \right] + N(k)
\end{equation}
where $P_{\rm dust}(k)$, $P_{\rm cib}(k)$, $P_{\rm source}(k)$ and $N(k)$ 
are respectively the power spectrum of the 
dust emission, of the CIB, of individually detected point sources and of the noise. 
The factor $\gamma(k)$ represents the instrumental function.

\subsection{Noise power spectrum}
\label{noise}

One of the main limitation of the component separation to
the power spectrum is the estimate of the noise contribution.
For ISSA maps, this contribution can be estimated accurately
as they are the combination of up to three HCON maps.
Indeed the noise power spectrum of the HCON maps, and therefore of the ISSA map,
can be estimated by subtracting two HCON maps of the same region. 
An example of the difference between two HCON maps for a typical low brightness ISSA map
is shown in Fig.~\ref{noise_ps} (top-right). The characteristic stripping of the IRAS data
is seen in this difference map.
The distribution of difference values is symmetric (when there is no strong point sources) 
and well fitted by a sum of two Gaussian functions (see Fig.~\ref{noise_ps}, bottom-right).

As the distribution of the difference values is the sum of two Gaussian functions with
an equivalent width $\sigma_{\rm diff}$ and assuming that the noise in the IRAS survey
is stationary to a good approximation (as tested below), we can conclude that the noise
of an individual HCON can also be characterized by an equivalent width:
\beq
\sigma_{\rm hcon} = \frac{\sigma_{\rm diff}}{\sqrt{2}}.
\eeq
Furthermore, the noise of the ISSA map, built from $n$ HCON maps, is also Gaussian
distributed with a width of:
\beq
\sigma_{\rm issa} = \frac{\sigma_{\rm hcon}}{\sqrt{n}}.
\eeq
Therefore, the noise level of the ISSA map is given by:
\beq
\sigma_{\rm issa} = \frac{\sigma_{\rm diff}}{\sqrt{n} \sqrt{2}}.
\eeq
For the example given in Fig.~\ref{noise_ps}, the noise level is $\sigma_{issa}=0.048$ MJy/sr.

In general three HCON maps were averaged but parts of them may be undefined.
Therefore,  for each ISSA map we looked at the three corresponding HCON maps and 
built a mask $n(x,y)$ that gives the number of defined values that were averaged at each positions.
Then, the noise map of a given ISSA map is estimated by subtracting 
two HCON maps and divide the result by $\sqrt{2}\sqrt{n(x,y)}$.

The power spectra of the ISSA map and of its noise map are shown in Fig.~\ref{noise_ps}(bottom-left). 
Both power spectra meet at small scales ($k \sim 0.2$ pixel$^{-1}$)
where the signal is noise dominated. The noise map is characterized by a $\sim k^{-0.8}$ power spectrum.
We have investigated the variations of the statistical properties of the noise
on ISSA maps where three complete HCONs exists. By looking at the three possible
difference maps built from three HCONs, we found that 
the shape of the power spectrum
of the noise is rather constant with time and that the absolute level of the noise varies 
by less than 15\% from one HCON to the other. 
This is equally true at 60 and 100 \ump.

\begin{figure}[!ht]
\hspace{-0.7cm}
\includegraphics[width=\linewidth]{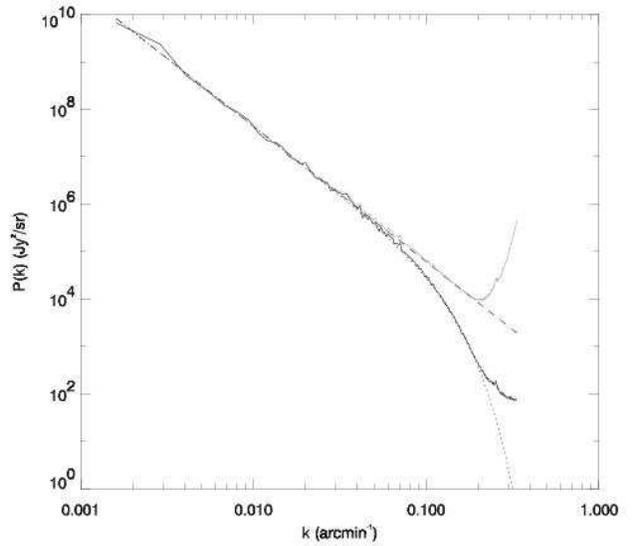}
\caption{\label{fit_psf} Power spectrum (black solid line) of a typical bright ISSA field with no strong point sources.
The cirrus contribution to the power spectrum has been fitted by a 
power law ($P(k) \propto k^{-2.86}$ for $k<0.015$ arcmin$^{-1}$
- dash line). The dotted line is the cirrus fit multiplied by a Gaussian PSF of $\sigma=0.07$ arcmin$^{-1}$.
The grey line is the power spectrum of the ISSA map (black solid line) divided by the Gaussian PSF.}
\end{figure}

\subsection{Point spread function}
\label{secPSF}

The multiplicative factor $\gamma(k)$ in Eq.~\ref{pk} represents 
the effective instrumental filter function of the system
(telescope, focal plane assembly, sky scanning and map making). 
For some ISSA maps the instrumental function can be estimated directly from the power
spectrum of the map. This is the case for maps where the brightness fluctuations are dominated by the cirrus
emission and where the power spectrum is simply:
\begin{equation}
P(k) = \gamma(k) \times P_{\rm dust}(k).
\end{equation}
Following \cite{gautier92}, the power spectrum of the cirrus dust emission follows a power law
($P_{\rm dust}(k)=Ak^\beta$). This behavior has been confirmed
through a detailed study of \hi 21 cm emission by \cite{miville-deschenes2002a}. These studies show
that although the power spectrum slope may vary from region to region
by $\sim$20 $\%$ there is no systematic break at high spatial frequencies.
On the basis of these studies, we thus conclude that the power law
representation for the cirrus power spectrum does not introduce any systematic effect.

To estimate $\gamma(k)$ directly on the power spectrum of ISSA maps, we have selected 20 relatively bright
regions (mean brightness greater than 8 MJy/sr)  where the noise and CIB contributions are negligible.
To make sure that the fluctuations are dominated by the cirrus emission, point sources were filtered
out using a median filtering. An example of the power spectrum of such 
an ISSA map is shown in Fig.~\ref{fit_psf}.
We found that the effective IRAS instrumental filter function in Fourier space is well describe 
by a Gaussian function both at 60 and 100 \ump. 
To determine the width of the instrumental function, 
the power spectrum of the 20 fields were fitted by the following equation:
\beq
P(k) = \exp\left(\frac{-k^2}{2\sigma_k^2} \right) \times Ak^\beta
\eeq
which represents the cirrus power law multiplied by the Gaussian instrumental function.
We have found $\sigma_k=0.073 \pm 0.005$ at 60 \um and  $\sigma_k=0.065 \pm 0.005$
at 100 \ump. In the real space, this corresponds to a Gaussian instrumental function
of $\sigma=1.5\pm0.1$ arcmin at 60 \um and $\sigma=1.7\pm0.1$ arcmin at 100 \ump.
\footnote{ This is compatible with what we found by fitting a Gaussian function on point
sources.}

The cutoff seen in Fig.~\ref{fit_psf} at frequencies larger than 0.05 arcmin$^{-1}$ 
is where the effect of the instrumental filter function is expected. 
When the power spectrum is divided by the Gaussian instrumental function (grey line)
one sees that this modeling does not stand
for scales smaller than 4.5 arcminutes ($k > 0.22$ arcmin$^{-1}$). 
This is due to the effective resolution of the IRAS data.
The power spectrum has to be cut at the Nyquist frequency
corresponding to the 1.5' pixel ISSA map.
However, the effective resolution of the IRAS data is 
about 3.8' and 4.25' respectively at 60 and 100 $\mu$m. For that reason,
all the power spectrum at frequency above 0.1 arcmin$^{-1}$, which corresponds
to the Nyquist frequency of the effective resolution, has to be taken
with cautious.

\section{Separation of the Galactic and extra-Galactic components}

\label{cib}

In Figs.~\ref{pscib1} to \ref{pscib_fin} we present
the power spectrum of the 60 and 100 \um ISSA maps of our twelve fields (black line).
For all these fields, the power spectrum of the noise has been subtracted
and the result has been divided by the PSF $\gamma(k)$.
At large scale, all the power spectra follow a power law typical of cirrus emission 
\cite[]{gautier92}. But at $k \sim 0.01$ arcmin$^{-1}$ we notice 
a break in the spectra and an excess from the cirrus power law at 
small scales. This power excess is observed for all our fields, at 60 and 100 \ump.

\subsection{Contribution of strong point sources}

The power excess observed at small scales reveals the
presence of a component with a flatter power spectrum, 
like what would be expected from noise or randomly distributed point sources. 
On the one hand, at this stage, a noise contribution is unlikely as it has been well estimated and
removed accordingly.
On the other hand, in fields with such low cirrus emission, several extra-galactic point
sources were detected. In each of our fields, between 200 and 300 point sources are
listed in the IRAS Point Source Catalog (PSC), the vast majority of them being extra-galactic objects.
We remove the contribution of all point sources with flux greater than 1 Jy
at 100 $\mu$m (and the same sources at 60 $\mu$m).
The cut at 1 Jy is chosen such that the PSC above this limit will be nearly complete.
This gives a well define separation between the sources contributing to the unresolved CIB
and the resolved sources. This is necessary for any quantitative use of our results.
We choose this cut in flux as low as possible to remove the contribution of the 
nearby and resolved galaxies that dominate the power spectrum at scales $k > 0.01$ arcmin$^{-1}$ 
(for those galaxies, the spatial distribution is indistinguishable from 
a Poissonian distribution). Moreover, according to recent simulations, the cut of 1 Jy at 100~$\mu$m 
almost corresponds to the flux where the non-euclidean part appears in the number counts.
To remove the contribution of strong point sources ($I_{100 \mu m} \geq 1$ Jy) to the power spectrum,
they were filtered out from the ISSA map. 
For each ISSA map, we have extracted the point sources of the IRAS 
Point Source Catalog with a flux greater or equal to 1 Jy.
Then we have applied a median filtering to the ISSA map to estimate the background level and, 
at each point source position,we have removed the points in a $12\times12$ window
that were more than 4 times the noise level ($\sigma_{\rm issa}$) above the background.
Finally, the missing points were replaced by a bilinear interpolation.

The power spectrum of the filtered ISSA maps is the blue line in Figs~\ref{pscib1} to \ref{pscib_fin}.
A power excess at small scales ($k>0.02$ arcmin$^{-1}$) is still apparent on all power spectra, at 60 and 100 \ump. 
This excess is likely to be the signature of the unresolved cosmic far-infrared background.
To characterize it, the contribution of the cirrus emission has to be removed.

\subsection{Contribution of the cirrus emission}
As we do not have any independent tracers of the cirrus emission at the IRAS scales
for the whole fields, the only way to derive the cirrus contribution is to use
the statistical properties of their spatial distribution.
Following \cite{gautier92} and \cite{miville-deschenes2002a} the power spectrum of interstellar dust emission 
is well describe by a power law. To remove this contribution to the power spectrum
of the filtered ISSA maps, we have fitted a power law on the large scale part
of the power spectrum of the filtered maps. The best compromise between the
number of point on which to fit the power law and the contamination from the CIB
was to fit on $k < 0.02$ arcmin$^{-1}$. The point corresponding to the largest scale
was not used for the fitting as it suffers from a large statistical error. 
The result of the fitting, and its associated uncertainty, 
are shown for each ISSA map in Figs~\ref{pscib1}
to \ref{pscib_fin}. The uncertainty on the slope reflects the dispersion of
the points around the fit.
The red line in these figures is the result of the subtraction of
the cirrus power law from the power spectrum of the filtered map (blue line);
it is the signature of the CIB.

\begin{figure*}
\setlength{\unitlength}{1.cm}
\begin{picture}(0,8)
\put(0,0){\includegraphics[width=8cm]{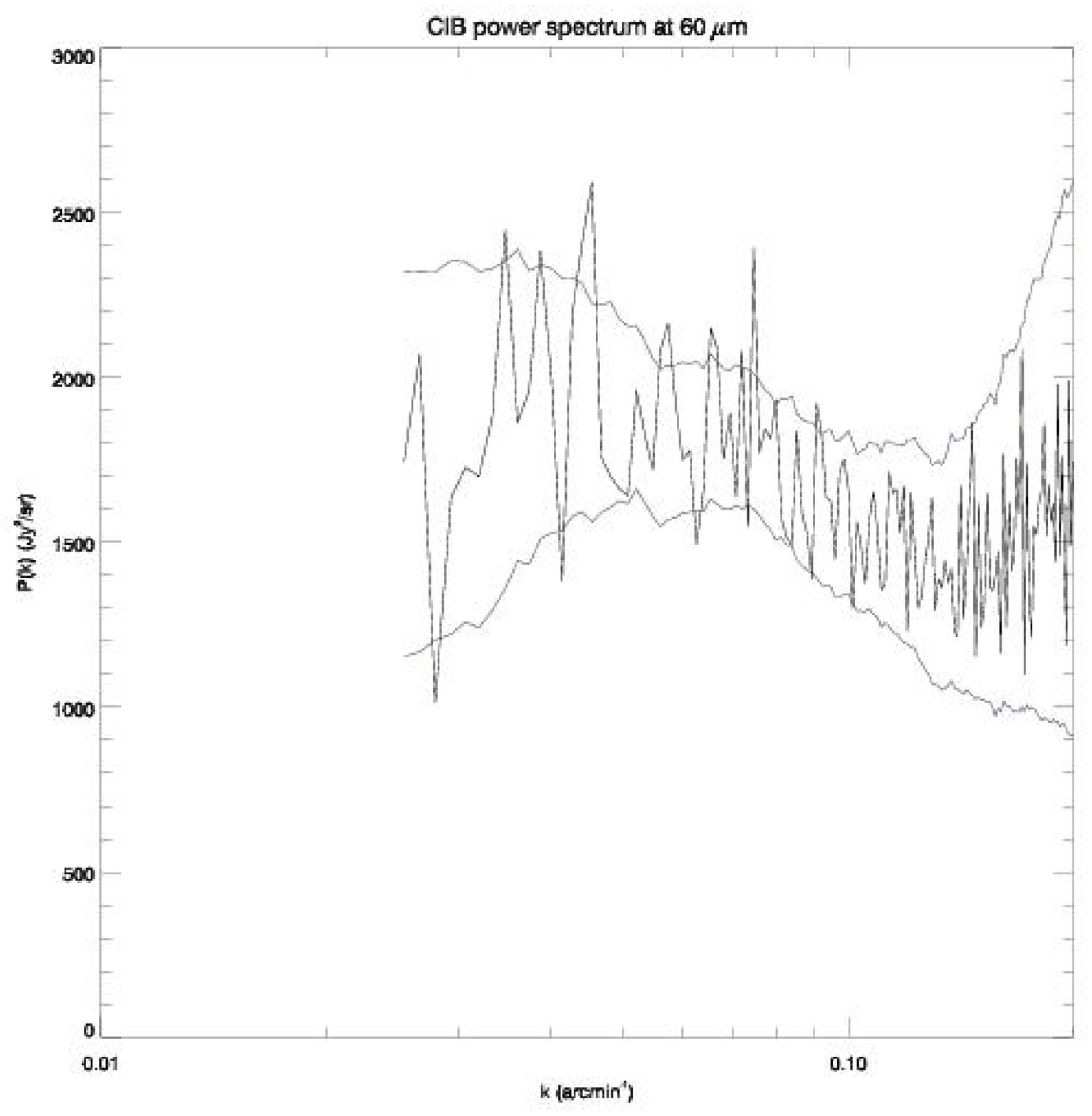}}
\put(9.5,0){\includegraphics[width=8cm]{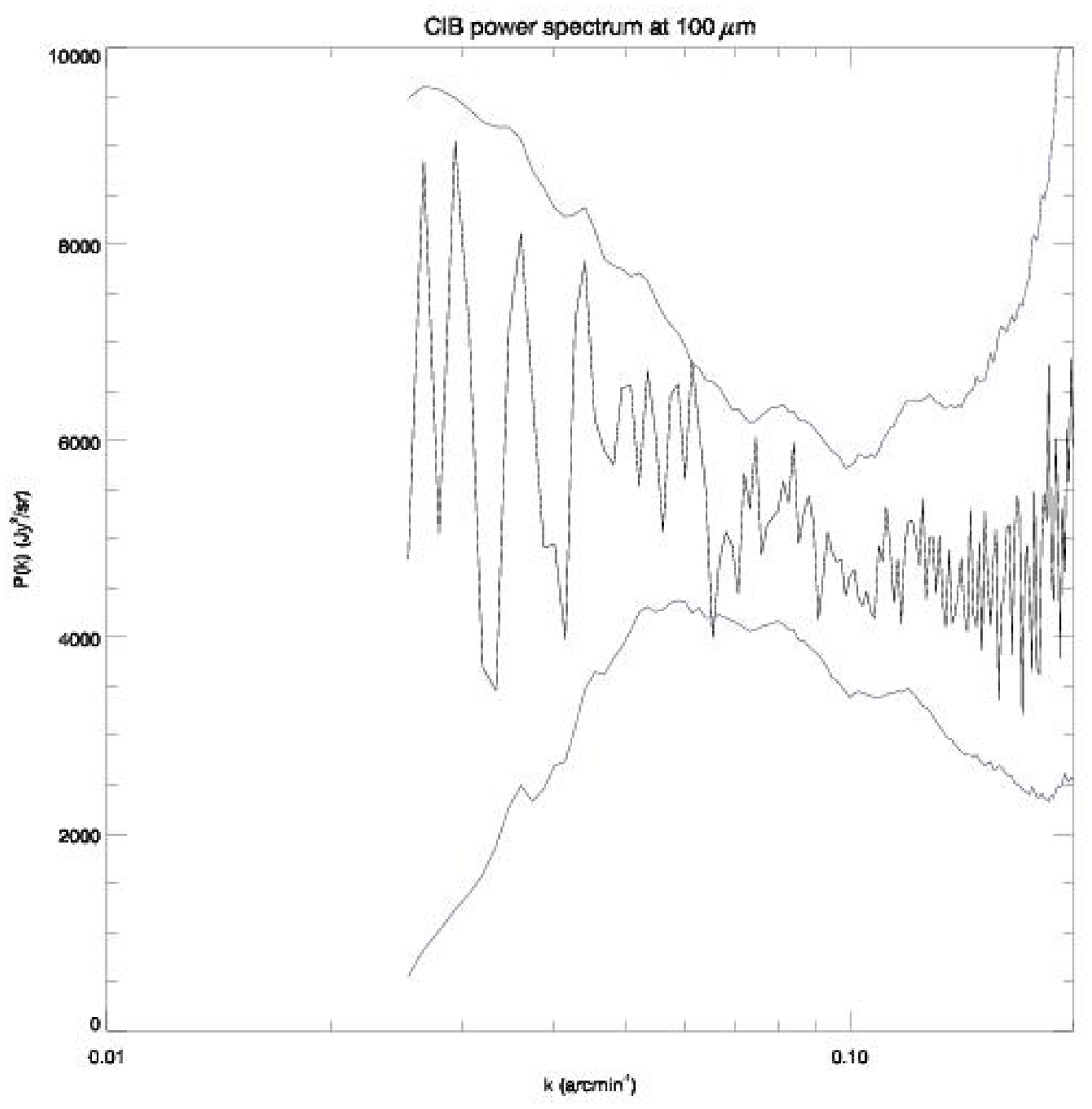}}
\end{picture}
\caption{\label{pscib_avg} CIB contributions to the 60 (left) and 100 \um (right) power spectra. The black line is
the average of the CIB contributions for the 12 fields. The blue lines indicates the uncertainty
on the CIB contribution, computed from the uncertainty on the cirrus slope and on the PSF width.
}
\end{figure*}

\subsection{Contribution of the cosmic far-infrared background}

The average of the 12 CIB contributions is shown in Fig.~\ref{pscib_avg} (black lines).
The blue lines in this figure indicate the 1$\sigma$ uncertainty on the determination, including
the PSF and the estimate of the cirrus contribution errors.
The 1$\sigma$ uncertainty is very close to the rms variations of the 12 averaged values.
The dispersion between the fields is thus dominated by the PSF and cirrus contribution
errors, we do not see any significant variation from field to field.
Moreover, it is important to note that the detected excess at small scales
is uncorrelated with the cirrus emission (the excess disappears when
the cirrus contribution increases). Therefore, our fields being
spread over the sky, the excess at small scale
is compatible with isotropic properties. This is an other
argument in favor of an extra-galactic origin
for the excess. \\
We can note on Fig. A.2-9 that the power spectrum
of the resolved IRAS sources is varying by factor up to 10
from field to field. The CIB (resolved and unresolved) fluctuations
are dominated by the brightest sources in ISSA maps.
The number of such sources is quite small, leading to a high
cosmic variance, and thus large variation from field to field. 
On the contrary, for the unresolved CIB fluctuations,
as sources above 1 Jy (about 300 per field) are removed,
the cosmic variance is lower than 6$\%$, which is negligible
with respect to the 1$\sigma$ uncertainty.\\
The CIB is detected over one decade in scales, from 5 to 50 arcminutes. 
The power spectrum of the CIB at 60 and 100 \um is compatible with 
a Poissonian distribution at spatial frequencies between 0.025 and 0.2 arcmin${-1}$.
The fluctuation level is $\sim 1.6\times10^{3}$ Jy$^2$/sr and 
$\sim 5.8\times10^{3}$ Jy$^2$/sr at 60 and 100 \um respectively.

\section{Discussion}

\label{discussion}

\subsection{CIB fluctuations at 60 and 100 \um}

We have shown here that the power spectrum 
of high latitude fields at 60 and 100 \um is characterized by a
break at small scales (near 0.02 arcmin$^{-1}$). 
As indicated by \cite{gautier92}, the power spectrum of the dust emission
is usually well described by a power law proportional to $k^{-3 \pm0.2}$. 
A more detailed study of the statistical properties of the
interstellar cirrus HI 21 cm emission has been carried out by \cite{miville-deschenes99b}.
In this analysis it is shown that
there are limited variations of the spectral index of the power law
from field to field but, what is most important for the present work, 
the power spectrum of cirrus emission for scales smaller 
than $12.5^\circ$ is always characterized by a single power law 
with no break. Therefore, it is unlikely that the power excess
observed here at small scales is of interstellar origin.

We could also wonder if this break is of instrumental origin.
It was shown by \cite{wheelock93} that the response of the IRAS
detectors are affected by memory effects. This produces variations of
the detector response as a function of scale. This effect is more
important at small scales (under a few tens of arcminutes) but \cite{wheelock93} have shown 
that the amplitude of the fluctuations at these 
scales were underestimated. This effect will thus produce
a drop of the power spectrum at small scale and cannot explain the
power excess observed here.

On the other hand, instrumental noise could produce such an excess in the power spectrum.
But, as the IRAS ISSA maps result from redundant individual observations, we were able to estimate
the contribution of the noise to the power spectrum.
We are aware that our estimate of the noise rely on the
fact that the individual HCONs are independant. This is not perfectly 
true as, in the construction of the HCONs, an offset was added to each scan to minimize the difference between different 
observations of the same position. Therefore the noise level estimated by 
subtracting HCONs may be underestimated at the scale of a scan, which is a few degrees. 
But at this angular scale the signal is completely dominated by the cirrus emission, even in 
the low brightness regions selected for our analysis. At the scale of a few arcminutes where
the CIB is detected, the noise contribution to the power spectrum has been removed accurately.

\begin{figure*}
\hspace{-0.7cm}
\includegraphics[width=\linewidth]{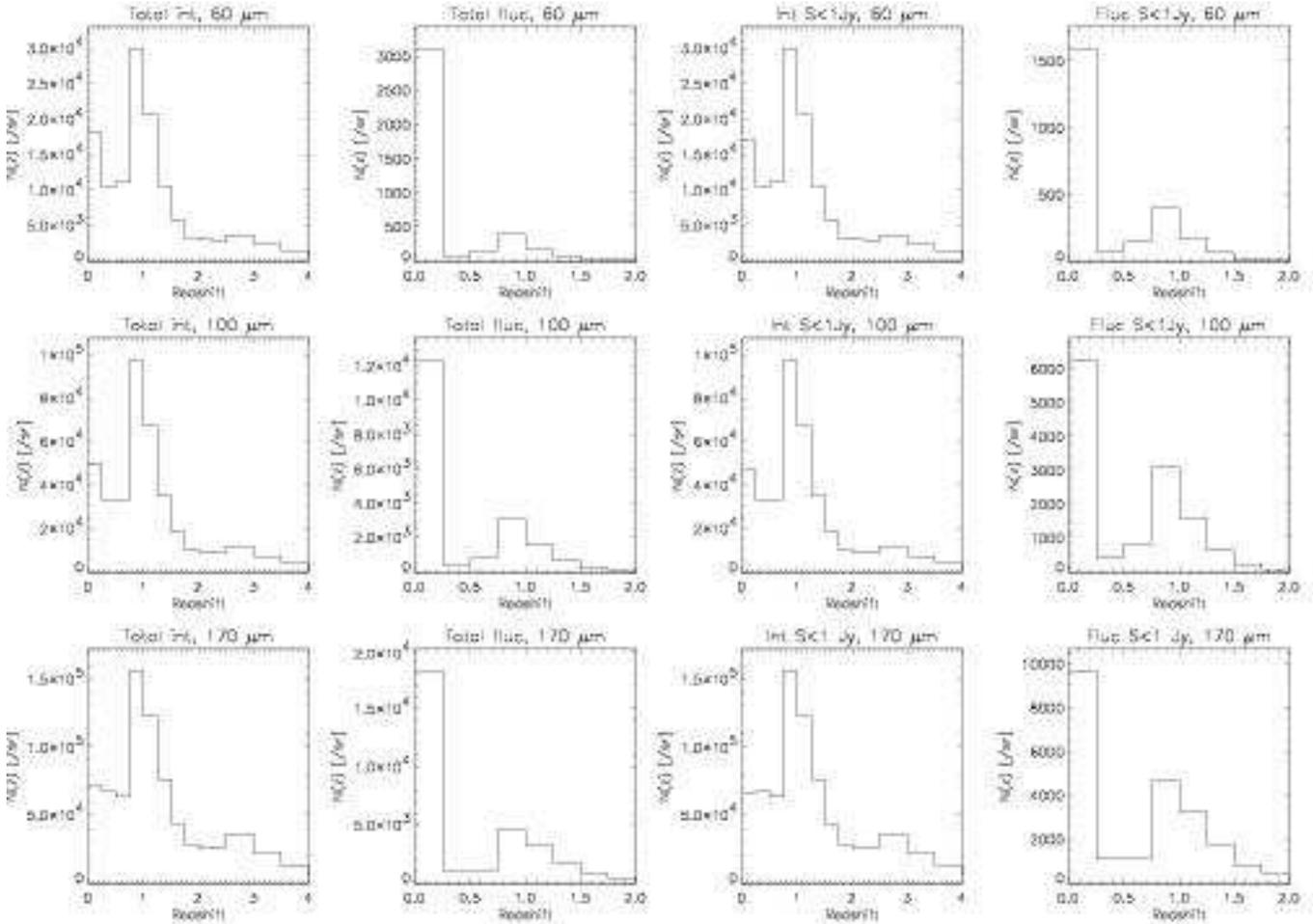}
\caption{\label{z-distrib} Redshift distribution of the sources
making the CIB and the fluctuations at 60, 100 and 170 \ump.
The first two-left panels are for the total contributions,
the last two-right panels, for sources with flux below 1 Jy
\cite[]{lagache2002}}
\end{figure*}

In fact it appears that most of the power excess can be attributed to the
numerous extra-galactic point sources that are present in such a low
cirrus emission field. When the strong ($I_{100 \mu m} > 1$ Jy) point sources are removed from the
ISSA maps, we recover a power spectrum typical of cirrus emission 
at low spatial frequencies but with
still a power excess at small scales (k$> 0.02$ arcmin$^{-1}$) that can be attributed
to the unresolved cosmic infrared background.
Moreover, the residue has homogeneous properties over the sky, 
consistent with CIB.\\

\cite{knox2001} computed the expected power spectrum of the CIB at several
frequencies ($\nu \le$ 1060 GHz), exploiting the far-IR volume emissivity derived from the
count models of \cite{guiderdoni98} and assuming a bias $b=3$, constant with
redshift.  They concluded that the clustering-induced fluctuations
can match those of the CMB at $\ell\lesssim$300.  
They also predict a shape of the CIB power spectrum peaking
around scales of 1-3 degrees. This broad maximum, if present, is at the limit of our frequency range
where the noise is becoming large,
making the detection of the clustering very difficult.
The power spectra of the CIB at 60 and 100 \um are compatible with a
Poissonian distribution with levels $\sim 1.6\times10^{3}$ Jy$^2$/sr and 
$\sim 5.8\times10^{3}$ Jy$^2$/sr respectively.

\subsection{CIB intensity and anisotropy amplitudes color ratio}

The CIB rms fluctuations in the IRAS maps 
corresponding to the white noise power spectra are: 
\begin{equation}
\sigma^{2}= \int P_{\rm cib}(k) 2 \pi k dk \quad Jy^2/sr^2
\end{equation}
giving $\sigma$=0.048 MJy/sr and $\sigma$=0.09 MJy/sr at 60 and 100 \um respectively.
As shown by \cite{gispert2000}, the CIB at different wavelengths is dominated by sources at
different redshifts, larger wavelengths being dominated by more distant
sources. The same applies to the fluctuations: lower frequencies probe 
higher redshifts (e.g. \cite{knox2001})
The fluctuations at 60 \um are dominated by nearby bright objects. 
When we remove these objects, the residual fluctuations
are quite low. At 100 \ump, the contribution of higher redshift 
objects is increasing, leading to higher level of
residual fluctuations.  
Therefore, the ratio of the 60 to 100 \um fluctuation is qualitatively
consistent with what is expected.

This is illustrated more quantitatively in Fig.\ref{z-distrib}
on the panels showing the redshift distribution of sources
contributing to the CIB intensity and the fluctuations \cite[]{lagache2002}.
It is clear from these figures that the z-distribution of the fluctuations
is bimodal, with one contribution at redshift lower than 0.25
and the other one centered at redshift around 1. 
For all sources the ratio of nearby to moderate-redshift
source contribution to the fluctuations is equal to 
3.8, 1.8 and 1.3 at 60, 100, 170 \um
respectively, illustrating that fluctuations at larger wavelengths 
are dominated by more distant sources.
When the brightest sources are removed the ratio 
of nearby to moderate-redshift
contribution becomes equal to 1.7, 1 and 0.74
at 60, 100, 170 \ump. In this case, at 100 \um
the contribution to the fluctuations
of nearby and moderate-redshift sources
is the same, becoming lower at higher wavelength.
At 60 \ump, the fluctuations are still dominated by
the nearby objects.

For the three wavelengths, the CIB is mainly due to sources
at redshift around 1.  A detailed analysis of
the CIB fluctuations at 100 and 170 \um (which is beyond the scope of this paper)
will give information on the distribution of sources at z$\sim$1 making
the bulk of the CIB. 
This is particularly true at 170 \um
where sources with flux lower than 4$\sigma$=135 mJy
can be removed \cite[]{dole2001}, leading fluctuations highly 
dominated by the moderate-z sources.

We can compute the ratio of CIB fluctuations to intensity 
($R_{\lambda} = \frac{\sigma_{\lambda}}{I_{\lambda}}$)
at 100 \um and compare it with the previous determination at 170 \ump.
To compute R$_{100}$ and R$_{170}$, we use:
\begin{itemize}
\item a CIB intensity at 100 $\mu$m of 0.5 MJy/sr \cite[]{renault2001}, 
leading to R$_{100}$ = 0.18
\item a CIB intensity at 170 \um of 1 MJy/sr \cite{lagache2001}.
For the fluctuations, we assume that a cut of 1 Jy at 60 and 100 \um
corresponds to the same cut at 170 \ump. We obtain rms fluctuations of
0.12 MJy/sr (corresponding to $\sim$11000 Jy$^2$/sr, \cite[]{puget2002}).
This gives R$_{170}$ = 0.12
\end{itemize}
The ratio R is decreasing between 100 and 170 \ump, as
expected. At 60 \ump, the minimal hypothesis is to consider that
R$_{60}$ is equal to R$_{100}$=0.18 which gives an upper limit of 0.27 MJy/sr
on the CIB intensity. An illustrative logarithmic extrapolation of R out to 60 \um
gives R$_{60}$=0.27 leading to an estimate of the CIB intensity 
at 60 \um of 0.18 MJy/sr, which is significantly smaller
than the previous determination of \cite{finkbeiner2000}
of 0.56 MJy/sr.

\subsection{Implication for the component separation}

This work suggests that the high latitude IRAS maps, in the lowest cirrus regions,
cannot be used as a tracer of the interstellar extinction structure as proposed
by \cite[]{schlegel98}. In fact it should be noted that it is only above
an intensity of order of 10 MJy/sr at 100 \um that the CIB fluctuations
are lower than the cirrus contribution at the smallest scales (k$\sim$0.2 arcmin$^{-1}$).

Present and future CMB observations, above 100 GHz, with high sensitivity bolometers
need to remove foreground contributions (cirrus and CIB fluctuations). The CIB spectrum
being significantly ``colder'' than the cirrus spectrum (${I_{\rm cirrus}100\mu m} / {I_{\rm cirrus}1mm} \sim$ 30;
${I_{\rm CIB}100\mu m} / {I_{\rm CIB} 1mm} \sim$ 5), the relative contribution
of the CIB will increase with wavelength. It is thus expected that at 1 mm
the range in l space dominated by the CIB will be much more extended than sees at
100 \ump. This question will be dealt in a forthcoming paper.

\section*{Acknowledgements}

{\em The Fond FCAR du Qu\'ebec provided funds to support this research project.}

\newpage

\begin{appendix}

\section{Power spectrum computation method}

\label{cross_fft}

\begin{figure}[!ht]
\includegraphics[width=\linewidth]{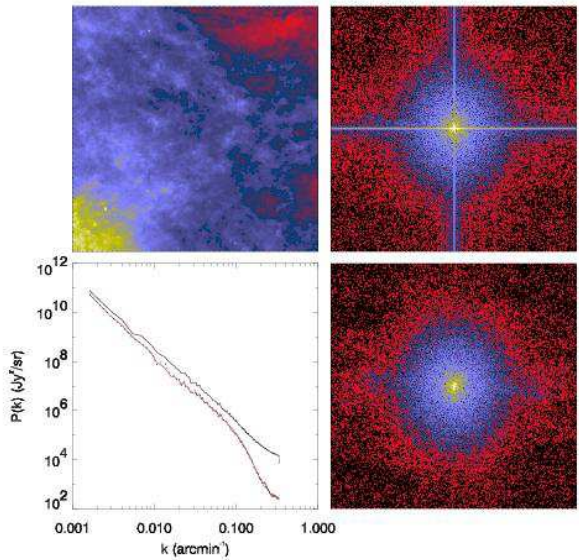}
\caption{\label{amplitude_fft} {\bf Upper left:} Typical ISSA map
with a large scale gradient. {\bf Upper right:} Amplitude of the 
Fourier Transform of the image on the left. The cross seen here is 
produced by the left-right and top-bottom
flux difference in the image. {\bf Lower right:} Amplitude of the Fourier Transform
of the image apodized with a cosine function to minimize the left-right and top-bottom
discontinuities. {\bf Lower left:} Power spectrum of the original image (solid line)
and of the apodized image (red line). The power spectrum computed with the Median method
is also shown (dotted line). It follows almost perfectly the power spectrum of
the apodized image.}
\end{figure}

The power spectrum of an image $f(x,y)$ of Fourier Transform 
$\tilde{f}(k_x,k_y)$ is computed from the
amplitude $A(k_x, k_y)$ defined as
\beq
A(k_x,k_y) = \tilde{f}(k_x,k_y)\tilde{f}^\star(k_x,k_y) = | \tilde{f}(k_x,k_y) |^2.
\eeq
The power spectrum $P(k)$ is the average of $A(k_x,k_y)$ on annulus of constant
$k = \sqrt{k_x^2 + k_y^2}$.
The amplitude image of a typical ISSA map is shown in Fig~\ref{amplitude_fft} (top-right).
In this figure, the frequency $k=0$ is a the center of the amplitude image. The amplitude
decreases with increasing value of $k$ but one also notices the presence
of a cross that reveals an increase of the fluctuation level in the
horizontal and vertical directions. This cross is caused by the Fast Fourier
Transform algorithm that makes an ``infinite pavement'' with the image prior to compute
the Fourier Transform. As the ISSA images are not periodic objects, this operation
produces discontinuities where the left (top) meets the right (bottom) side
of the image. These horizontal and vertical discontinuities are responsible
for the cross seen in the amplitude image. The main problem with this cross
is that it increases artificially the power when the average over $k$ is done.
Furthermore, the power increase is not constant as a function of $k$ 
which modify the slope of the spectrum as well as the normalization.

The usual method to get rid of this effect is to apodize the image so that
the left (top) and right (bottom) sides of the image have a similar flux level.
Another method which do not modify the image but acts on the amplitude image itself is to 
compute $P(k)$ by taking the {\it median} value at constant $k$. For $\sim$85\% of
the ISSA maps we have inspected, these two methods give very 
similar results. But when the cross is very strong, 
the power at small scales is still overestimated by the ``median'' method.
Furthermore, at small $k$ values, there are less points to average and
a large fraction of them are affected by the cross effect; 
at these frequencies, the median value may not be significant.
For all these reasons, we have adopted the ``apodization'' method.

\newpage

\begin{figure*}[!ht]
\includegraphics[height=0.8\linewidth,angle=90,draft=false]{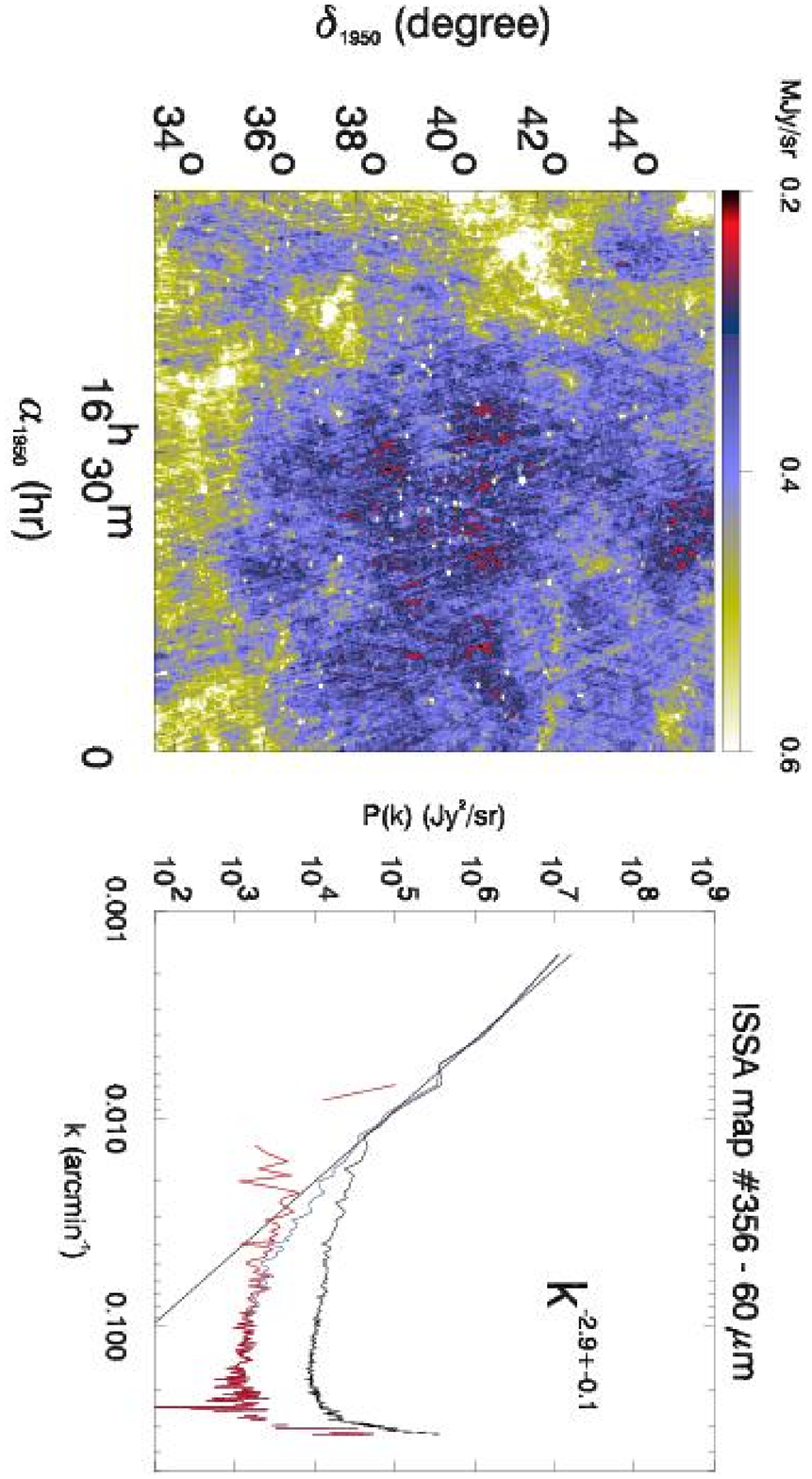}
\caption{\label{pscib1} Power spectrum of a typical 60 \um ISSA maps.
The black curve is the power spectrum of the raw ISSA map from which the power spectrum of
the noise has been subtracted. The blue curve is the power spectrum 
(also noise subtracted) of the map from which point sources stronger than 1 Jy in the
IRAS point sources catalog were removed. A power law fit of the blue curve 
(for $k < 0.02$ arcmin$^{-1}$) is also shown. The red curve is the difference 
between the blue curve and the power law fit.}
\end{figure*}

\begin{figure*}[!ht]
\includegraphics[height=0.8\linewidth,angle=90,draft=false]{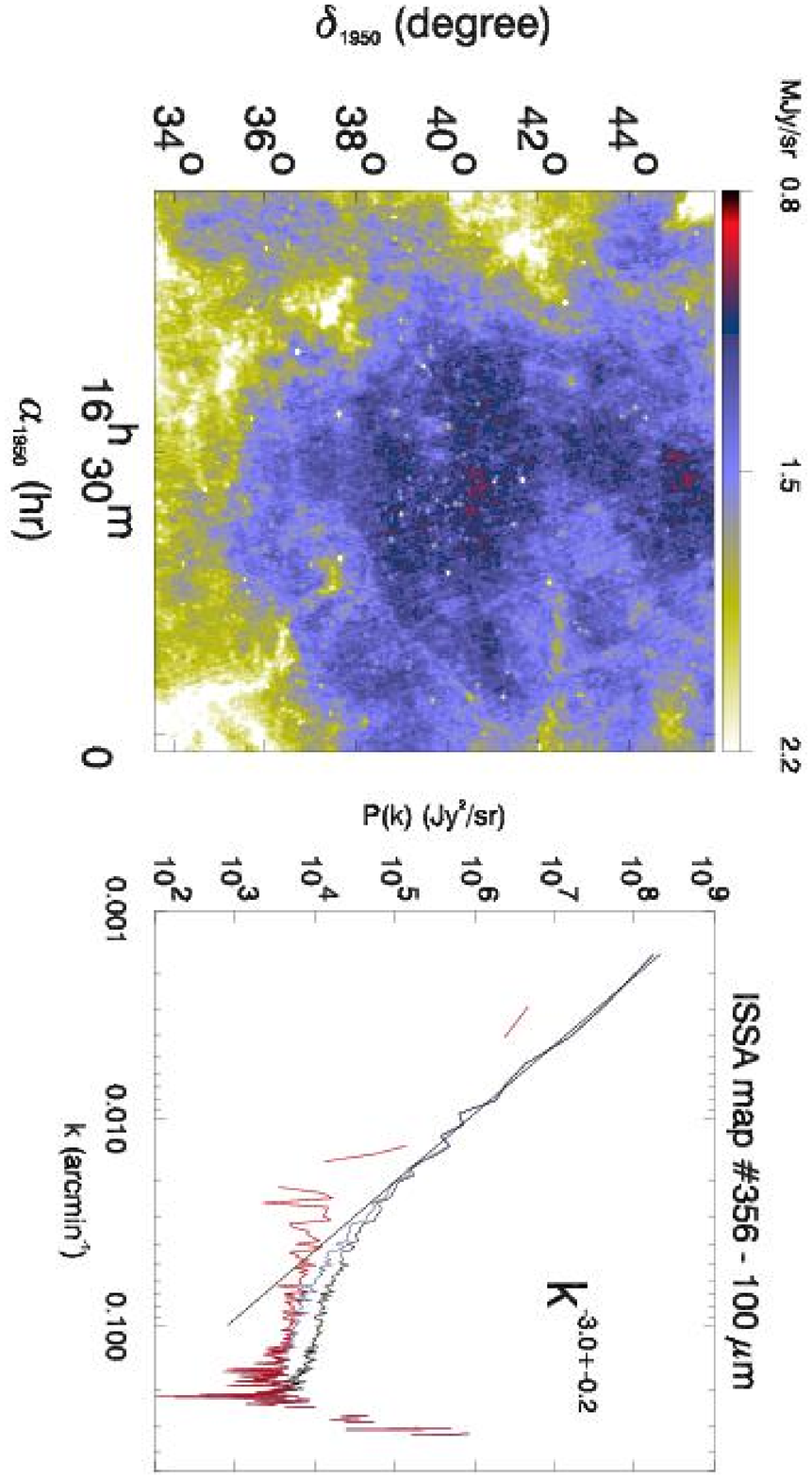}
\caption{\label{pscib_fin} Power spectrum of a typical 100 \um ISSA maps.
The black curve is the
power spectrum of the raw ISSA map from which the power spectrum of
the noise has been subtracted. The blue curve is the power spectrum 
(also noise subtracted) of the map from which point sources stronger than 1 Jy in the
IRAS point sources catalog were removed. A power law fit of the blue curve 
(for $k < 0.02$ arcmin$^{-1}$) is also shown. The red curve is the difference 
between the blue curve and the power law fit.}
\end{figure*}

\end{appendix}

\end{document}